\documentclass[a4paper, 12pt, oneside]{article}
\usepackage{amssymb}
\usepackage{indentfirst}   % русский стиль: отступ первого абзаца раздела
\begin{document}
\newcommand{\Dx}{\Delta q}
\newcommand{\drm}{\mathrm d}
\newcommand{\ep}{\epsilon}
\newcommand{\qarr}{\stackrel{\cal Q}{\longrightarrow}}
\newcommand{\lc}{L^2 (\Sg_3 (t);\ \bf C;\ b^3 (t){\sqrt \omega (t)} d^3 q)}
\newcommand{\vp}{\varphi}
\newcommand{\1}{{\bf\hat 1}}
\newcommand{\Oc}{O\left(c^{-2(L+1)}\right)}
\newcommand{\im}{{\rm i}} % should be substituted
                          %by \newcommand{\im}{{\rm i}}  in LaTex2e
\newcommand{\Sg}{\Sigma}
\newcommand{\bgt}{\bigotimes}
\newcommand{\ptl}{\partial}
\newcommand{\Sche}{ Schr\"odinger equation\ }
\newcommand{\Schr}{Schr\"odinger representation\ }
\newcommand{\eu}{$ E_{1,3} $}

\newcommand{\euf}{E_{1,3}}
\newcommand{\rif}{V_{1,3}}
\newcommand{\ri}{$V_{1,3}$ }
\newcommand{\ov}{\overline}
\newcommand{\stc}{\stackrel}
\newcommand{\defst}{\stackrel{\mathrm{def}}{=}}
\newcommand{\qstc}{\stackrel{\cal Q}{\longrightarrow}}
\newcommand{\h}{\hbar}
\newcommand{\beq}{\begin{equation}}
\newcommand{\nde}{\end{equation}}
\newcommand{\beqa}{\begin{eqnarray}}
\newcommand{\ndea}{\end{eqnarray}}
\newcommand{\rin}{$V_{1,n}$}
\newcommand{\rinf}{V_{1,n}}
\newcommand{\rn}{$V_n$}
\newcommand{\rnf}{V_n}
\newcommand{\al}{\alpha}
\newcommand{\be}{\beta}
\newcommand{\ga}{\gamma}
\newcommand{\om}{\omega}
\newcommand{\Sch}{Schr\"{o}dinger \ }
\newcommand{\Schs}{Schr\"{o}dinger's \ }
 \newcommand{\lcn}{$L^2 (\rnf; {\mathbb C}; \sqrt\omega d^n q)$ }
\begin{titlepage}
\title{On Geometric Paradox in   Quantization of  Natural  Dynamical Systems }
%\begin{titlepage}
  \author{E.~A.~Tagirov\\
\small {Bogoliubov Laboratory of Theoretical Physics,}\\ {\small Joint Institute for Nuclear Research,
   Dubna, 141980, Russian Federation}}
 \date{}
\end{titlepage}
\maketitle
\begin{abstract}The initial \Sch variational approach (1926)  to quantization  of the natural Hamilton mechanics   in  $2n$-dimensional phase  space  is  revised  in the  modern paradigm of quantum mechanics in  application  to   the  system the  Hamilton function of  which is a positive-definite  quadratic  form   of the $n$ momenta with the  coefficients  depending  on the  canonically conjugate coordinates in the generic Riemannian configuration space $V_n$.  The quantum  Hamilonian thus  obtained  has  a  paradoxical  potential  term depending on  choice of  coordinates in  $V_n$, which was discovered first  by B. DeWitt in 1952   in the  framework of  canonical  quantization  of the  system by  a  particular ordering of  the  basic  operators of observables  of  momenta  and  coordinates.
% (a  localizationof the  system in    $V_n$) .
 It is  shown that the   \Sch approach   in the  standard  paradigm  of  quantum  mechanics   determines uniquely  the ordering selected by DeWitt  among  a  continuum of other possibilties to  determine the quantum  Hamiltonian.

    Two  conceptually  important   particular classes of  observables of   localization in $V_n$  
are  considered   in detail. It is  shown  that, in general,  the  quantum-mechanical potential does not vanish even in  the Euclidean configuration space $V_n$ except  the case when  Cartesian coordinates are taken  as the observables of localization of the  system. It is  noted also that ,  in the quasi-classical approach to the quantization considered by DeWitt in  1957,   the quantum Green function (propagator) is also   non-unique and depends on the choice of a line in $V_n$ connecting these points.  All three formalisms have  the  same local asymtotics of  the  quantum Hamiltonian   if the normal Riemannian cordinates are used, at least implicitly, to  localize  the  system in  $V_n$ .
   \end{abstract}
%%%How wonderful that we are faced with a paradox. Now we have a hope for promotion!
\newpage
\begin{quote}\flushright\emph{How wonderful that we have met with a paradox.\\ Now we have some hope of making progress.\\ Niels Bohr}
\end{quote}

\section{Preface.}	 

A little-known, but ,at least, theoretically existing paradoxical phenomenon of the \emph{quantum-mechanical potential} (it will further be denoted  as QMP)  arises  generically  in   quantization of a  particular class of conservative  Hamilton systems  which are   called \emph{the natural  systems}.  They   are the systems whose  the (classical)  Hamiltonians functions $ H^\mathrm{(nat)}(q,p)$  are   the (positive-definite) inhomogeneous  quadratic   forms of  momenta $p_a   , \  \  a, b, ... = 1, 2 ,.., n $ with   coefficients   $\om_{ab}(q) \equiv \om_{ba}(q) $ and an external potential $ V^\mathrm{(ext)} (q)$ of which are sufficiently smooth   functions of coordinates $q^a$\ \ in an $n$-dimensional  configuration space $V_n$, that is
\beq
              H^\mathrm{(nat)}(q,p) = \frac1{2m} \omega^{ab}(q)p_a p_b + V^\mathrm{(ext)} (q), \
 \label{Hnat} 
\nde
where $m$  is a constant of dimension  of  mass,   ${q^a,  p_b}$  are canonically cojugate  coordinates    on  the  phase space $\mathcal {P}_{2n} \equiv T^*V_n$  which  has  thus a  symplectic  structure. The functions  $\omega^{ab}(q)$ and $ V^\mathrm{(ext)} (q)$  are supposed  to be   tensor fields on $V_n$ of   types  $\left({\textstyle\begin{array}{c}2 \\ 0\end{array}}\right)$ and 
$\left({\textstyle\begin{array}{c}0 \\ 0\end{array}}\right) \mathrm{(i.e.\ a\  scalar)}$   correspondingly
w.r.t. the arbitrary sufficiently smooth transformations of $q^a\in V_n$. Then,   $V_n$ is endowed  by a  Riemannian structure    with the metric  form  
 \beq
 \drm s_\mathrm{(\om)}^2 = \omega_{ab}(q)\drm q^a \drm q^b , \label{s1}
       \nde
where  $\omega_{ab}$ of  $V_n$ is determined as usual: $\om_{ac}\om^{cb} = {\delta_a}^b $. 
 %%%The functions  $f(q, p) \in     \\ !!!!
\\
In the  present  paper, the  so  called heyuristical level of  mathematics traditional in the theoretical physics is adopted. 
    
\section{Canonical quantization   of  natural  systems}

Let $ {\bf s}_{2n}$ be  an appropriate subalgebra  of the Poisson algebra
of real functions $ f \in C^\infty({\cal P}_{2n})$ which  are often called Hamiltonians in the mathematically rigorous texts.  
 In  the present    paper oriented  to  physics,  they will be called \emph{classical observables} and the     term (classical) Hamiltonian will be kept for  $ H^\mathrm{(nat)}(q,p)$ in  view  of  its staightforward  relation  to  the natural  Hamilton  mechanics'.   
According  to   \cite{Abr}, pp 425 - 434, existence on $V_n$  of  the  natural  measure 
$\om^{\frac 12}\mathrm{d}q^1...\mathrm{d}q^n , \om $ being the determinant of  the matrix 
$\|\om_{ab}\|$,   allows   to  construct   a quantum  counterpart to  the  generic natural  system by 
  the  linear map   
\beq
{\mathcal Q}: \ {\bf s}_{2n} \ni f
\qarr \hat f\quad (\mbox{operators in a pre-Hilbert space}\ {\cal H}), \nonumber
\nde
satisfying the following conditions:

 (Q1) $ \quad  1\ \qarr {\bf\hat 1} $\ (the identity operator in $\cal H$);

(Q2) $\quad  \{f, g\}_\h \ \qarr \
 \im\h^{-1} [\hat f, \hat g] \defst
\im\h^{-1}   (\hat f \hat g - \hat g \hat f)$ \quad
where $\{f, g\}_\h \equiv  \{f, g\}_0 + O(\h) $ is an antisymmetric
bilinear functional of $f$ and  $g$ and  $\{f, g\}_0 \equiv \{f, g\} $ is
the Poisson bracket  in ${\cal P}_{2n}$;

(Q3) $\quad \ \hat{\ov f} \qarr  (\hat f)^\dagger $
(the Hermitean conjugation of  $\hat f$ with respect to the scalar
product in $\cal H$ );

 (Q4)   a complete set of classical observables\footnote{As in  the  classical as well as in  the  quantum-mechanical  context,  the  term  "observable" means  that, basically,  a \emph{classical} physical procedure  exists,         at least, speculative. In  particular,  the set of observables $ \{q^a\}$ fixes a localization   of  the   system in  $V_n$.}  (maximal Abelian subalgebra)
 $f^{(1)}  , ..., f^{(n)}: \   f^{(a)}\in {\bf s}_{2n},$
is mapped to a complete set (in the sense by Dirac \cite{Dir},  
Chapter.III) of commuting operators $\hat f^{(1)}, ..., \hat f^{(n)}$.\\

  %% Stop
If  \emph{a paricular system of canonically conjugate coordinates $ q^a,  p_b $ is fixed} in 
${\bf s}_{2n} \equiv T^*V_n$ and the  coordinates  ${q^a}$ in $V_n$  are taken  as the complete  set  of  the classical obsrvables  of localisation  of  the  system  and operators
 $\hat q^ {(a)} \defst q^a \cdot  {\bf\hat 1}$  in  
${\mathcal H} \sim L^2 (\Sg; {\Bbb C};  \om^{\frac12} d^n q)$ are   their  quantum counterparts  under condition
(Q3), then   the  condition  (Q2)    will  be  fullfilled   for a map ${\mathcal Q}$ of the  classical  observables of   canonically conjugate  coordinates (\emph{the Darboux coordinates}) $q^a, p_b$    on $T^*V_n $ to  the symmetric operators (\emph{a configuration space representation of $n$-dimensional  Heisenberg algebra}).
\beq
q^a \ \qarr \ \hat q^a \defst q^a \cdot {\bf\hat 1}, \ 
p_b \ \qarr \ \hat p_b \defst  -\im \h \left (\frac{\ptl}{\ptl q^b} + \frac14\frac{\ptl}{\ptl q^b}\ln \om \right ). \label{qp}
\nde
The quantization  map ${\mathcal Q}$ of  the  metric tensor  $\om (q)$ can be  be defined  following the  Von Neumann rule  (\cite{Neu}, p. 313)  for functions
of commuting  operators  $\hat A_1, ... , \hat A_N$:
\beq
f(A_1,..., A_N ) \qarr   \hat f \defst f (\hat A_1, ... , \hat A_N).
\label{neu}
\nde
 Thus, the components of the  metric  tensor on $V_n$  is also  considered and quantized  as an observable:
\beq
\om_{ab} (q^1,...,q^n) \qarr \hat\om_{ab}(q^1,...,q^n)
 \defst \om_{ab }(q^1,..., q^n) \cdot {\bf\hat 1}. \label{om}
\nde
In 1952, B.~C.~DeWitt  \cite{DW1}  had taken the  following Hermitean  product of  these  operators as the quantization map $\qarr$ of the  generic  natural  Hamilton  function $H^\mathrm{(nat)}(q, p)$  :      
\beq
    H^\mathrm{(nat)}(q,p) \qarr \hat H^\mathrm{(DW)}(\hat q, \hat p) \defst \frac1{2m} \hat p_a \om^{ab}(\hat q)\hat p_b +    V^\mathrm{(ext)}(\hat q). \label{HDW}
\nde   
Then, in terms of  the  representations  (\ref{qp}), 
  (\ref{om}) (\emph{the  \Sch representation)} as differential operators  on $ L^2 (\Sg; {\Bbb C};  \om^{\frac12} \mathrm{d}^n q)$ 
\beq
 \hat H^\mathrm{(DW)} \equiv  -\frac{\h^2}{2m}\Delta^{(\om)}(q) 
 + V^\mathrm{(qm;DW)}(q) \cdot \hat{\bf 1} + V^\mathrm{(ext)}(q)\cdot \hat{\bf 1}, \label{HDW2}
\nde
where 
\beq  
\Delta^{(\om)}(q) \defst 
\om^{-\frac12}\frac{\ptl}{\ptl q^a}\left(\om^{\frac12} \om^{ab} \frac{\ptl}{\ptl q^a}\right) 
\equiv\om^{ab} \nabla^{(\om)}_a \nabla^{(\om)}_b   \label{Lap}
\nde
is the Laplace-Beltrami operator for $ V_n$,  $\nabla^{(\om)}_a$ is the  metric connection (covariant  derivative )  in $V_n$, and   
\beq V^\mathrm{(qm;DW)}(q) \defst 
  - \frac{\hbar^2}{2m}\ \om^{-\frac14}\frac{\ptl}{\ptl q^a}(\om^{ab}\frac{\ptl}{\ptl q^b} \om^{\frac14})\qquad\qquad\qquad \label{Vqm}
\nde
 is just   \emph{the DeWitt version of QMP.}
DeWitt    perceived this potential term as an  extraordinary  and  unsatisfactory result because 
$V^\mathrm{(qm; DW)}(q)$  \emph{is generically not invariant w.r.t. to  change  of   coordinates $q^a$},
contrary to  $H^\mathrm{(nat)}(q,p) $   and $\Delta^{(\om)}(q)$.   (Of course,   there  are  particular choices of  the  coordinates, e.g.,   such  that the coordinate  condition  
 $\frac{\ptl}{\ptl q^b} \om = 0$  is satisfied and thus $V^\mathrm{(qm; DW)}(q) = 0$ for that choice.)
 The presence of this potential in the quantum Hamiltonian obviously makes its spectrum dependent on the choice of the coordinates  $q^a$ that  is on  a  method of observation  of   localization of  the  system  in  $V_n$ in  the classical mechanics.  \emph{An amazing result not only for the 1950s but also for today's quantum physics,  isn't it?}. \\

 On  the  other  hand , in the quantum  field  theory, the  phenomenon of quantum  anomalies   is  well known 
when some conservation law  related  to  a symmetry that is  valid in the classical theory ceases to be satisfied if the quantum effects are correctly taken into account.  So  the  phenomenon of QMP may apparently be called  \emph{the  quantum  anomaly of covariance of   the  natural Hamilton  dynamics w.r.t. the  coordinate transformations  in  the  configuration  space $V_n$ (non-relativistic  general covariance)}.\\

As soon as $ \hat H^\mathrm{(DW)} $  is  determined, one  may  transform it,  at  least  locally,   to another curvilinear coordinates $\tilde q^a$  by substitution  $q^a =  q^a(\tilde q) $  as if  all terms in (\ref{HDW2}) were  invariant   w.r.t. to the substitution (i.e. are scalars) . Thus, 
 it   is very  important to distinguish the  phenomenon   of QMP generated  by  fixation  particular observables  $q^a$  from  the "technical" change of  an  expression  for  the  Hamilton operator   changed  by  transition  to  some new  coordinate  variables as  a appropriate tool to solve or to study the Schr\"{o}dinger equation with QMP .           \\
	
DeWitt  took $\hat H^\mathrm{(DW)}$   as  \emph{"the simplest method of symmetrising"}  of  the  product  of  operators $\hat q, \hat p, \hat \om $ . 
It is a matter of taste, however.   There is an one-parametric family  of  the hermitean  orderings of  the  product, which satisfy the Correspondence Principle:    
\beqa
  \hat H^\mathrm{(\nu)}
  &=&\frac{2-\nu}{8m} \om^{ab} (\hat q) \hat p_a \hat p_b
 + \frac\nu{4m} \hat p_a  \om^{ab} (\hat q) \hat p_b +  
 \frac{2-\nu}{8m}\hat p_i \hat p_b \om^{ab} (\hat q)  \label{Hnu}\\
  &= & \hat H^\mathrm{(Sch)} +  V^\mathrm{(qm;\nu)}(q) \cdot \hat{\mathbf 1} \label{Hnu1}\\
  V^\mathrm{(qm;\nu)}(q)&\equiv& V^\mathrm{(qm)}(q)
 + \frac{\h^2(\nu-2)}{8m}\ptl_a\ptl_b \om^{ab}. \qquad  \label{Vnu}
         \ndea
This is a typical ambiguity that arises not only in the canonical formalism of quantization  and constitutes one of the main problems for the transition from classical to quantum theory in  the  Hilbert-space-based formalisms.\\

Again,\emph{the potential  $V^\mathrm{(qm;\nu)}(q)$ depends generically  on  the choice  of  coordinates $q^a$  for any  value  of   parameter} $\nu$  except  the  case  when  $V_n$ is Euclidean and  Cartesian coordinates $x^i$  are taken  as $q^a$  in it.\\

 In 1957, DeWitt \cite{DW2} had  returned  to  the  problem  of  quantization in  the quasi-classical  formalism, which he  considered   as   leading   to  the  result invariant w.r.t.  the arbitrary  transformations of coordinates in  $V_n$.   However,  in  the  quasi-classical  approach,    these coordinates correspond to the "technical" changes of  coordinates mentioned above.  Dependence on the choice of "observation" is  included  into  a propagator (a two-point  Green  function).  The 
 quasi-classical approach and  its local coincidence with the result of  the above canonical approach will be  discussed in   Section  5 below.    But first I will state a strong argument  in favor of $\hat H^\mathrm{(DW)}$    arising  in the  Schrödinger variational approach to  quantization  \cite{Sch1}  after a  recent review of it  in  the modern paradigm  of  quantum  mechanics \cite{TAG1}. 

\section{The Schr\"{o}dinger variational  quantization of natural systems and its modernization}
It seems to be  a  little-known  fact that quantization of the generic natural system
was   considered first  by E.~Schrodinger in the third  of  his  five papers  submitted to "Annalen der Phyzik" in the  first  half of  1926, by which he had founded  the wave mechanics\cite{Sch2}. The main purpose of his study  was  \emph{to construct the quantum dynamics which  plays the same role w.r.t. the  classical  dynamics that  the wave theory  of light does w.r.t. the geometric  optics}. To  this  end, he had taken   the following  positive-definite functional  of  the    "wave functions"  
 $\psi(q)$ from the normalized  space $\mathcal{H}^{(r)}$ of  \emph{real}(!) functions      
 \beqa
  J^\mathrm{(Sch)}[\psi] &= & \int_{V_n} \omega^{\frac12}\,\drm^n q \,\left\{\frac{\hbar^2}{2m}
    \left(\ptl_a \psi\, \omega^{ab}\,\ptl_b \psi\right)
    + \psi^2 V^\mathrm{(ext)}(q)\right\},  \label{Phi}\\
  \psi(q) \in \mathcal{H}^{(r)},  &  &   \ptl_a\defst \frac{\ptl}{\ptl q^a} , \nonumber
    \ndea
  subordinated to  the  additional condition   
  \beq
   \int_{V_n} \omega^{\frac12} \drm^n q \,\psi^2 = 1 , 
%\quad \om  \stackrel{\mathrm{def}}{=} \mathrm{det}\|\om_{ab}\|,            
         \label{nrmr}
  \nde
Schr\"{o}dinger interpreted  $J$ as   the energy mean value of the \emph{real} field  $\psi (q)$ corrsponding to the  quantum state of the  natural system under consideration .  Then  the Euler-Lagrange equation  with  the right-hand term of  (\ref{Phi})  and accounting condition (\ref{nrmr}) with a Lagrange multiplier $E$   comes to  
 \beq
   \hat H^{(\mathrm{Sch})} \psi = E \psi , 
\quad  \hat H^{(\mathrm{Sch})}\defst
  -\frac{\h^2}{2m}\Delta^{(\om)} + V^\mathrm{(ext)}  ,     \label{Hscv}   
 \nde
\\

The  Schr\"{o}dinger Hamiltonian $\hat H^{(\mathrm{Sch})}$ looks  a very  satisfactory quantum  counterpart  to $ H^\mathrm{(nat)}(q,p)$ : it depends on the observables  of   localization $q^a$ in  the  configuration space $V_n$ only through the Laplace-Beltrami operator $\Delta^{(\om)}$ and the external potential $V^\mathrm{(ext)}(q)$, and \emph{is invariant w.r.t. to  choice  of  coordinates $q^a$}. In few weeks after publication of   Schr\"{o}dinger's work,  Eq.  (\ref{Hscv}) had been used  by F.~Reiche \cite{Reich} to  calculate  the spectrum of  a spherical top  considering  its  configuration  space as  the  three-dimensional  sphere  with   the identified  antipodal points. Later, this  result  was developed for calculation  of  spectra  of more  complex  molecules.\\ 

The point, however, is that  \Sch  considered the wave functions $\psi(q)$ as   real  physical  fields  forming  
\beq
                 {\mathcal H} \sim {\mathcal H}^{\mathrm(r)} \sim  L^2 (V_n; {\mathbb R};\sqrt \om\drm^n q).
\nde 
His last  paper  among  the mentioned basic ones \cite{Sch2} was devoted just  to the search for physical meaning of the wave function $\psi(q)  \in {\mathcal H}^{\mathrm(r)}$.\\

In the  modern QM, however, the space of  wave functions is  the {pre-Hilbert space of complex-valued functions}
\beq
      {\mathcal H} \sim {\mathcal H}^{\mathrm(c)} \sim  L^2 (V_n; {\mathbb C};\sqrt \om\drm^n q).
\nde 
The variational \Sch quantization has been  reconsidered in the modern paradigm of  non-relativistic quantum mechanics based on  ${\mathcal H}^{\mathrm(c)}$ in \cite{TAG1} .
It  unambiguosly lead  to the Hamiltonian $ \hat H^\mathrm{(DW)}(\hat q, \hat p)$ introduced  by DeWitt.\footnote{Apparently,   \Schs approach  to  quantization of  the  generic natural  systems  was not  known to DeWitt since he  never referred   to  it.}
Thus, QMP   $V^\mathrm{(qm)}(q)$ , or, equivalently,  $\nu =2$ in the  general  rule of  ordering and DeWitt's choice  (\ref{HDW})  of  the  quantum Hamiltonian are distinguished  unambiguously and \emph{the problem  of ordering  of the elements  of the Heisenberg algebra and the operator
 $\hat\om^{ab} $ in the Hamilton  operator is solved}. \\

Since the  QMP depends on  choice  of observables of   localization   $\hat q^a$ in  the  configuration  space $V_n$, consider two  importantant  particular  cases:  \emph{quasi-Cartesian} (or \emph{ normal Riemannian)  coordinates} and  \emph{small deformations   of the  Cartesian  coordinates} $x^i$ in $E_n$. \\

\section{Quasi-Cartesian, or normal  Riemannian, coordinates in $V_n$} 

These coordinates are denoted further as $y^{(a)}$. Let $q_0 \in V_n $ is an  origin  of the  coordinates $y^{(a)}(q)$  determined by  the components of the tangent  vector in $q_0$ to  the geodesic  line from  $q_0$ to $q$  in an orthonormal frame $\lambda_b^{(a)}$ at $q_0$  as follows  
 \beq 
y^{(a)}(q) \defst s(q, q_0)\left(\frac{\drm q^b}{\drm s}\right)\Biggm|_{q=q_0}\lambda_b^{(a)} ,        \label{y}
\nde
where $s(q, q_0)$ is the geodesic  distance between $q_0$ and $q$, see, e.g.,  \cite{Syn}, Ch.II, Sec.8 . \\

\emph {Geodesic lines have  no  intrinsic curvatures. Thus, the coordinates $y^{(a)}$ completely determined  by the Riemannian structure  of $V_n$.} In  a vicinity of the origin $q_0$     
\beq
\drm s_\mathrm{(\om)}^2  = \left(\delta_{(ij)} - \frac13  R^{(\om)}_{(ikjl)} (q_0) y^{(k)} y^{(l)} + 
O(s^3) \right)\drm y^{(i)} \drm y^{(j)} ,
\label{s3} 
\nde
where $R^{(\om)}_{abcd}$     is  the  Riemann--Christoffel curvature tensor of  $V_n$ determined so  that
$(\nabla^{(\om)}_a\nabla^{(\om)}_b - \nabla^{(\om)}_b\nabla^{(\om)}_a)f_c   = R^{(\om)d}_{abc}f_d $ for any twice    differentiable 1-form $f_c(q)$.
From here,  the asymptotic expression of  zero-order in  $s$ for  QMP, Eq. (\ref{Vqm}), is
 \beq
V^\mathrm{(qm;DW)}(y)= \frac{\h^2}{2m}\left(\frac{1}{6} R^{(\om)}(q_0) + O(s)\right) = 
 \frac{\h^2}{2m}\left(\frac{1}{6} R^{(\om)}(q) + O(s)\right)             \label{Ry} 
\nde
 $R^{(\om)}(q)$ is the \emph{Ricci scalar curvature} of $V_n$ .(In the  expression for $ V^\mathrm{(qm)}(y)$, it  is used that $R^{(\om)}(q_0)  = R^{(\om)}(y) +  O(s)$.) 
However,  it  should  not be missed that  Eq.(16)  is  a   result  of the  special   choice of  observation of    localization in    $V_n$.  Then,   the zero-order asymptotic quantum Hamiltonian  in  the 
time-independent Schr\"{o}dinger equation     is
\beqa
   \hat H^\mathrm{(DW)}(y) &=& -\frac{\h^2}{2m} \left(\left(\frac{\ptl}{\ptl y^{(1)}}\right)^2 
   + \left(\frac{\ptl}{\ptl y^{(2)}}\right)^2 + \dots +  \left(\frac{\ptl}{\ptl y^{(n)}}\right)^2  \right)  \nonumber\\
  & + &    V^\mathrm{(qm;DW)}(y) +   V^\mathrm{(ext)}(y) + O(s),  \label{Hmy}
 \ndea 

%% Thus, the zero-order asymptotic expression of QMP is
%%%%% ЗНАК ПЕРЕД R ВЫБРАН МИНУС ! КАК В "UNFINISHED...1" ПРОВЕРИТЬ
 %%\beq V^\mathrm{(qm)}(y)=- \frac{\h^2}{2m}\left(\frac{1}{6} R^{(\om)}(q_0) + O(s)\right) = 
 %%\frac{\h^2}{2m}\left(\frac{1}{6} R^{(\om)}(q) + O(s)\right)             \label{16R} 
%%$R^{(\om)}(q)$ is the \emph{Ricci scalar curvature} of $V_n$.\\
\emph{NB! The expression  for QMP is  valid  only for the canonical quantization  with the particular choice $q^a \equiv y^a  $ of the phase space   coordinates:} 
\beq
y^{(a)} \rightarrow
 \hat y^{(a)} \equiv  y^{(a)}\cdot \hat {\bf 1} ,  \   
p_{(b)} \rightarrow \hat p_{(b)}  \equiv   - \im \h\left(\frac{\ptl}{\ptl y^{(b)}} + O(s)\right).
\nde
After fixing thus the asymptotic method  of  observation  of   localization of  the  quantum  system in $V_n$,     one may  transform  quasi-Cartesian coordinates $y^{(a)}$ to arbitrary  curvilinear coordinates $\tilde q^a$  in  a small domain   at    the origin $q_0$ and  thus represent there the asymptotic  Hamiltonian as   
\beq
\hat H^\mathrm{(DW)}  \approx -  \frac{\h^2}{2m} \left(\Delta^{(\tilde\om)} +
 \frac16 R^{(\tilde\om)} (\tilde q)\right) +  \tilde V^\mathrm{(ext)}(\tilde q)   \label{H16R} 
 \nde
where $\Delta^{(\tilde\om}$ is  the Laplace-Beltrami operator  in  "arbitrary" coordinates $\tilde q^a$.  
 However, recall again that  Eq.(\ref{H16R}) is  a  result  of "technical" transformation   to coordinates 
$\tilde q^a$  from the coordinates  $y^a$  initially taken  for  quantization as observables  of   localization in $V_n$ , and that  $y^a$ are in  fact  two-point  functions determined  by  the  geodesic line  between   $q_0$ and $q$ .  

\section{QMP in  the Euclidean  configuration space}

 QMP is  generally non-zero even  in  the Euclidean $V_n$ if  the observables  $q^a$ are  not  Cartesian coordinates.  Denoting the latters  as  $x^a$, let us   calculate   the  first  order  approxmation of  QMP for  the  case  in which   $q^a = x^a + \ep f^a(x)$ where $\ep$ is  a  small  parameter and $ f^a(x)$ are   arbtrary sufficiently differentiable functions This  means  a  small deformation  of  the system of   Cartesian coordinates. 
 A little  algebra  gives  for  this  case 
\beq
           \hat H^\mathrm{(DW;\ep)}  = -  \frac{\h^2}{2m}
\left(\Delta^{(\om)} +  \frac{\ep\h^2}{4m} \Delta \mathrm{Tr}\|\ptl_a f^b\| + O (\ep^2)\right), 
\nde
%% check!
where the second summand is  the  asymptotic QMP  and $\Delta$  is  the  Laplace operator in the Cartesian  coordinates $x^a$(as well in $q^a$ in  the  approximation  under  consideration) .

%% It is  remarkable that,  at least in the accepted asymptotic  approximation,
%%QMP, as well  as  other  first  order  terms in $\ep$, vanish, if $f^a$ is  a  Killing  vector (the 
%% generator  of an  isometry  of $V_n$) .\\

This  particular  example  shows more clearly that   QMP is  tightly related  to  the  conceptual  problem of  mesurment in quantum  mechanics. Contrary to the  case   of Subsection 3.1, where  the    meаsurment  of the   localizationin  $V_n$  is determined in terms of  the interior Riemannian structure of  the natural  system, now  a  non-trivial QMP is generated by introduction  of  an  external  vector field slightly deforming  the Cartesian coordinates .  It may  have also a practical  meaning, since a real physical  observation  of   localization  can never be  precise  even the  Cartesian coordinates  are   taken  for  that. 
%%%% Stop
\section{Quasi-classical  quantization}

In 1957,  DeWitt \cite{DW2} had returned  to a thorough study  of  the relation between   (non-relativistic) classical and  quantum dynamics'   in   the  curved  configuration   space.  
Considering the  evolution in time   of the  wave function 
$\psi(q, t) \in L^2 (\rnf; {\mathbb C}; \omega^{1/2} \drm^n q) $ as determined, in  general, by a  propagator $ G(q, t| q', t')$:
\beq
\psi(q, t) 
 = \int_{V_n} \om^\frac12(q') \drm^n q' \, G(q, t| q', t')\, \psi(q', t'), \quad  q, q' \in V_n, \label{G}
 \nde
  DeWitt had  generalized  the  Pauli  construction  \cite{Pauli2} of $G(q, t| q', t')$   for a charge in an e.m. field in  the Cartesian coordinates in the flat  $V_n$ to the  case of  the generic natural  system: 
\beqa && G(q, t| q', t') = \nonumber\\
 &&\qquad =\omega^{-1/4} (q)\, D^{1/2} (q, t|q', t')\,\omega^{-1/4} (q')
 \, \exp\left(-\frac i\hbar  S(q, t|q',\, t') \right), \label{GDW}\\
 &&D (q,\, t|q', t' )\defst \det\left(- \frac{\ptl^2 S}{ \ptl {q}^i \ptl {q'}^j}\right)\quad
  \mbox{(the  Van Vleck  determinant)} \nonumber
\ndea
  and  $S(q,\, t|q',\, t')$ is  a  solution  of  the  Hamilton-Jacobi equation  for
 $H^{(\mathrm{ nat})} (q,p)$.  (In  fact, initially  he had included interaction  with an external   vector field  into the  Hamiltonian, but ,   for the sake of simplicity,  it will be omitted here as well as the  term with  the  external scalar potential $V^{\mathrm (ext)}$ as DeWitt had done in  the  final  part of  his  paper.)\\

Construction  (\ref{GDW})  is in fact  the postulate of quasi-classical (or, WKB)  quantisation .  Using the Hamilton-Jacobi equation for  the natural system DeWitt had obtained    that  the quasi-classical propagator 
$ G(q, t| q', t')$ 
  \emph{"nearly satisfies the \Sch equation"}:   
   \beq
 i \hbar \frac{\ptl}{\ptl t} G(q, t|q', t')=
 \left( - \frac{\h^2}{2m} \Delta^{\mathrm (\om)}  
 +  \widetilde{V}^{\mathrm{(qm)}}(q, t|q', t') \right)  G(q, t|q', t') \label{G1}
      \nde
  where
 \beqa
    \widetilde{V}^{\mathrm{(qm)}}(q, t|q', t')&\defst&  \frac{\h^2}{2m}\ 
\frac{\ptl_i \left(\om^{\frac12}(q) \ptl^i\left(\om^{-\frac14}(q) D^{\frac12}(q, t|q',t')\right)\right)}
{\om^{\frac14}(q)\, D^{\frac12}(q, t|q',t')}\label{Vtilde}
 \ndea
In fact, Eq.(\ref{G1})  is just  the \Sch equation with QMP $\widetilde{V}^{\mathrm{(qm)}}(q, t|q', t') $  for the  propagator  $G(q, t| q', t')$,  and this  QMP    is  evidently a two-point function  of $q$ and $q'$.  and, in general, is a functional of  the line  connecting them. 	So one  sees that $\widetilde{V}^{\mathrm{(qm)}}$    \emph{ only looks as a   scalar with  respect to transformations of the  coordinates  in  neighborhoods  of the  points $q^a$ and $q'^a$}. However,  one  should keep in  mind  that the two-point functions $ G(q,t|q',t')$ and $D(q,t|q',t')$ are  correspondingly \emph{a bi-scalar} and \emph{a bi-density of  the  weight $-1$} at   both  points\footnote{An  excellent  introduction to  the technique   of  bi-scalars,  bi-tensors    and  bi-densities is given  in  the  paper  \textit{Radiation  Damping in a Gravitational Field} \cite{DWB}
by  B.~DeWitt and R.~Brehme. However, their  main  conclusion is  not  correct  that   the general--relativistic Lorentz--Dirac equation  has  no term including  values of Riemann-Christoffel  curvature tensor  at   the point  of  localization and  thus  the  principle of  equivalence is fullfilled. This was corrected in  \cite{Hobbs} for the case of  the electric  charge and in \cite{Ilin} for the case of point-like charge of the  scalar field which  is  more  relevant to the context of  the  present paper. }.
In a more  general context,  they  depend on a choice of the bi-scalar $S(q,t|q',t')$ of  the  classical  action.    In  the simplest  case    under consideration, it determines
 the   geodesic  dynamics  in $V_n$  and therefore the  situation  is equivalent to  introduction  of  the  Riemannian  coordinates $y^a $ as basic observables   in  Section 4. The system of arbitrary  coordinates $\{q^a\}$  in  Eqs.(\ref{G1}), (\ref{Vtilde})  
plays the technical   role mentioned  there. Therefore, it is  not  surprising that,   from   Eq.(\ref{Vtilde}),  DeWitt comes  to the  following zero order asymptotics: 
 \beq
\widetilde{V}^{\mathrm{(qm)}}(q, t|q', t') = \frac{\h^2}{2m}\, \frac16 R^{(\om)}(q)+ 
o(q-q')+ o(t-t'), \label{Vqmt}
\nde
 that is  just to   $V^\mathrm{(qm)}(y)$  in Eq.(\ref{Ry}), Section 4. Thus, in  this  approximation,     DeWitt's quasiclassical  result  coinside  with  canonical  QMP   in the  version  of  ordering  of  the basic operators    determined  by  the modern \Sch  QMP
 $V^\mathrm{(qm)}(q) $, Eq.(\ref{Vqm}),  i.e.  with   $\nu=2$  chosen  in  Eq.(\ref{Vnu}).\\

The  quasi-classical  local asymptotic  by  DeWitt  had  been focused on the use  in  the    path integration  approach to   quantum mechanics.  In  this  aspect, it has  attracted  quite a lot of attention  and has  been applied  to more  general non-relativistic  mechanical systems, which demand  to  take  into  account the  asymptotic  terms   of higher orders  of  $q-q'$ in the quasi-classical propagators .   A large analytical  review of these  systems and   problems related  to  them in the  path integration  is given   by L.~V.~Prokhorov \cite{Prokh} in 1982.  He  repeatedly noted   ambiguities  appearing in  the formalism. In 
 paper  \cite{TAG1} ,  a uniquely determined  causal quantum propagator for  the generic natural system have been constructed     in the original Feynman  formulation of the path  integral, which supports again  the  result of  the  present  paper. But it should be the matter of  a  special publication.\\   

\section{Conclusions}  
The  quantum-mecahnical  potential (QMP), arising  at quantization  of  the generic classical natural system,  violates the symmetry (general  covariance) of  the  original  classical Hamiltonian under generic  change of the  observable of  localization  of the system in  $V_n$. (So,  the  Darboux  coordinates in  the  phase  space are determined,  too). This phenomenon may  be interpreted as \emph{a quantum  anomaly  of  general covariance}.  It  may to take  place  in a much  wider range of  dynamical systems where   geometry of the  phase  space comes into  play.
% An example of this kind within the framework of geometric quantization is given in \cite{DKal}.
 From this  point  of  view,  the subject of the present paper  has a great potential for development and improvement. 

Returning  to the  natural  systems and the  formalisms of  quantization considered in  the  present paper, namely  DeWitt's canonical  and quasi-classical approaches , and the  modernized  \Sch variational approach, the  following conclusions  need to be emphasized.\\
1. When $R\neq 0$, the modernized  \Sch  varational quantization  leads  not  only  to  the same QMP which the canonical approach by DeWitt   gives  but   distinguishes     it unambiguously  ( requires $\nu =2$ in Eq.(\ref{Hnu1})) among infinitely  many other possibilites  of  ordering  of the  basic operators  $\hat q, \hat p, \hat \om $ in  the quantum  Hamilton  operator $\hat H^{(\nu)}$. \\
2.   The varational  and  canonical quantizations give rise to  a closed expression for 
QMP contrary to the quasi-classical  approach which is  asymptotic and, in fact,  is implicitly related   to the  choice of   normal Riemannian (quasi-Cartesian) coordinates.   \\   
3. The both former  approaches   lead  to the  generically non-vanishing QMP even  in  the Euclidean
  configuration   space $E_n$ except the  case when the  Cartesian  coordinates  are  taken for localization of the  quantum system. This effect can  be understood from  the  physical point  of  view:
any  coordinate  line,  except the Cartesian ones,   has  intrinsic  curvatures which  can  be   considered  as manifestation(s) of  external  forces  on  trajectories   of   the  thought  instruments for observation   of localizaion  of the  system.
% The  quantum  Hamiltonian 
  \\    
4. The  most intriguing  result  is  that, when the   normal Riemannian coordinates $y^{(a)}$ are taken in $V_n$, it  follows   
\beq
       V^{(\mathrm{qm}} (y)  =   -  \frac{\h^2}{2m} \frac16 R(y) + O(s),                                 \label{conf}
 \nde
 from all  three considered formalisms of quantizationas as the first non-vanishing term   of  the local asymptotics of QMP  in  a  vicinity of  the  origin  of  the coordinates. (Comparison the following  terms  of  the  asymptotic deserves calculation.) 
 In  its geometric  meaning,  this is just the term that  R.Penrose \cite{Pen} introduced into the  \emph{massless} general-relativistic Klein-Gordon  equation (the general- relativistic  generalization  of the \Sch equation)  for it  to  become conformally covariant. (Penrose has considered,   of course, only
	the \emph{four-dimensional} space-time, i.e. $n=3$.)  However, as it  was shown in \cite{ChT},   a much more comprehensive  study  of  the role  of  conformal covariance  in the  quantum field theory of a scalar field with  $m^2 \geqq 0 $   and arbitrary space dimension  $n$,    the conformal covariance of the  equation  in  the particular case of  $m^2=0$  requires that    the  additional term  must be 
\beq
           -  \frac{\h^2}{2m} \frac{n-1}{4n} R , 
 \nde
  that  is  it coincdes with  (\ref{conf}) on the  appropriate  space sections of the globally-static  space-time   only   if $n = 3$. Since  the early 1970s  the  general-relativistic  scalar field  equation with  this additional term (exact, not  asymptotic) has been  called later  as  the equation  with conformal  coupling (the term Penrose-Chernikov-Tagirov equation  can also be met in  the literature) and  widely applied  in  the  inflationary cosmology.   However, a mystery is that  the asymptotic QMP (\ref{conf}) arises in  the  non-relativistic  quantum  mechanics \emph{without any visible connection with  general relativity  and  conformal  symmetry} and with the  same  coefficient $1/6$  in  front  of  the  Ricci scalar curvature $R$ of the Riemannian configuration space $V_n$ i.e. for  any  space dimension $n$. Thus,  non-relativistic quantum  mechanics distinguishes  the dimension of real space  $n=3$  among all thought Lorentzian possibilities.    \\

I would like to thank Dr. N.A. Tyurin for  a  mathematical  consultation.   


\begin{thebibliography}{99}
 %\bibitem{Got} M.~J.~Gotay,  \emph{Obstruction to Quantization}  in:\emph{Mechanics: From Theory %%to Computation}, J.Nonlinear Science Editors, pp. 271–- 316 (Springer, New York, 2000)\\	
\bibitem{Abr} R.~Abraham and J.~R.~Marsden, \emph{Foundations of Mechanics}, p. 421,  
The Bendjamin/Cummings Publishing Company, Inc., 1978.\\
\bibitem{DW1} B.~S.~DeWitt, Phys.Rev. \textbf{85} (1952), p.653.\\
\bibitem{Sch1} E.~Schr\"odinger, Ann.d.Physik , \textbf{79} p.734; (1926).\\
\bibitem{Sch2} E.~Schr\"odinger, Ann.d.Physik , \textbf{79}, p.361; ibid. \textbf{79}, p.489
 ibid. \textbf{79}, p.734; ibid. \textbf{80},p. 437; ibid. \textbf{81}, p.109. (1926).\\
\bibitem{TAG1} E.~A.~Tagirov,  Gravitation and Cosmology, \textbf{19} (2013),  p.1 .\\
\bibitem{Syn} J.~L.~Synge,  \emph{Relativity: The General Theory},  North Holland Publising Company, 1960. \\
\bibitem{Pauli}   W.~Pauli, \emph{Wellenmechanik}, in: \emph{Handbuch der Physik}, 1933, Band 24, I, p.120.\\
\bibitem{Dir}  P.~A.~M.~Dirac, \emph {Principles of Quantum Mechanics},
Cambridge University Press, Cambridge, 1958.\\
\bibitem{Neu} J.~ Von Neumann,, \emph{Mathematical Foundations of Quantum
Mechanics}, Princeton University Press, Princeton, 1955 .\\
\bibitem{DW2} B.~S.~DeWitt, Rev.Mod.Phys.\textbf{29} (1957), p.377.\\
\bibitem{Reich} F.~Reiche, Z. Phys. , \textbf{39} (1926), p.444.\\
\bibitem{Pauli2} W.~Pauli,  \textit{Feldquantisierung. Lecture notes.} (1950--1951), Zurich.
\bibitem{DWB} B.~S.~DeWitt,  R.~W.~Brehme,  Ann.Phys.(N.Y.), \textbf{9}  (1960), p.220. 
\bibitem{Hobbs} J.~M.~Hobbs, Ann.Phys.(N.Y.), \textbf{41}  (1968), p.191.
\bibitem{Ilin}    S.~B.~Il'in,  E.~A.~Tagirov, Theoretical and Mathematical Physics, \textbf{37} (1978), p. 885.
\bibitem{Prokh}  L.~V.~Prokhorov,  Soviet Journal of Particles and Nuclei,  \textbf{13}(5) (1982),
 p.456-482. 
%\bibitem{DKal}, Reports on Mathematical Physics, \textbf{43}  (1999), p.147
\bibitem{Pen} R. Penrose, \emph{Conformal  Treatment of Infifnity} in: \emph{Relativity, Groups and Topology}, B.S. DeWitt ed.,
Gordon and Breach, London, 1964, p.255.
%%\bibitem{Son} S. Sonego, \textit{Journ. Math. Phys.}a {\bf 51} (2010) , 092502.
\bibitem{ChT} N.A. Chernikov and E.A. Tagirov,  {\sl Annales de l'Institute
Henry Poincar\`e}, {\bf A9} (1968), p.109.
\end{thebibliography}
\end{document}